\documentclass[10pt, conference]{IEEEtran}
\ifCLASSINFOpdf
\else
\fi
\usepackage{subfigure}
\usepackage{graphicx}
\usepackage{hyperref}
\usepackage{authblk}
\usepackage[none]{hyphenat}

\begin{document}
\title{A Comparative Study of Containers and Virtual Machines in Big Data Environment}


\author[1]{Qi Zhang}
\author[2]{Ling Liu}
\author[2]{Calton Pu}
\author[3]{Qiwei Dou}
\author[3]{Liren Wu}
\author[3]{Wei Zhou}
\affil[1]{IBM Thomas J. Watson Research, New York, USA}
\affil[2]{College of Computing, Georgia Institute of Technology, Georgia, USA}
\affil[3]{Department of Computer Science, Yunnan University, Yunnan, China}


%Telephone: (800) 555--1212, Fax: (888) 555--1212}


\maketitle

\begin{abstract}
Container technique is gaining increasing attention in recent years and has become an alternative to traditional virtual machines. Some of the primary motivations for the enterprise to adopt the container technology include its convenience to encapsulate and deploy applications, lightweight operations, as well as efficiency and flexibility in resources sharing. However, there still lacks an in-depth and systematic comparison study on how big data applications, such as Spark jobs, perform between a container environment and a virtual machine environment. In this paper, by running various Spark applications with different configurations, we evaluate the two environments from many interesting aspects, such as how convenient the execution environment can be set up, what are makespans of different workloads running in each setup, how efficient the hardware resources, such as CPU and memory, are utilized, and how well each environment can scale. The results show that compared with virtual machines, containers provide a more easy-to-deploy and scalable environment for big data workloads. The research work in this paper can help practitioners and researchers to make more informed decisions on tuning their cloud environment and configuring the big data applications, so as to achieve better performance and higher resources utilization.
\end{abstract}


% \fi
\IEEEpeerreviewmaketitle

\section{Introduction}
The expanding and flourishing of IoT (Internet of things) has brought us into the era of big data. It is no doubt that cloud computing is being widely adopted for big data processing. To provide an elastic and on demand resource sharing, different kinds of commercial clouds, such as public cloud, private cloud, hybrid cloud, and etc., have been established, where virtual machines are used as building blocks of the cloud infrastructure to gain higher hardware resources utilization while preserving performance isolation among different computing instances. 

Despite its advantages, virtual machine based cloud also faces many challenges when running big data workloads. One example is its weak resilience. Although multiple virtual machines can share a set of hardware, the resources in one virtual machine cannot be easily shifted to another. Although many smart ideas have been proposed to enable efficient resource management for VMs \cite{peng2012vdn, knauerhase2004dynamic, xiao2013dynamic}, resource sharing in virtual machine environment remains to be a challenging issue \cite{singh2016survey}, and it is still not unusually to observe application performance degradation due to not being able to handle peak resources demands, even when there exist free resources \cite{zhang2017memflex, zhang2016iballoon}. Another example is the poor reproducibility for scientific research when the workloads are moved from one cloud environment to the other \cite{boettiger2015introduction}. Even though the workloads are the same, their dependent softwares could be slightly different, which leads to inconsistent results.

Recently, container-based techniques, such as Docker\cite{docker}, OpenVZ \cite{openvz}, and LXC(Linux Containers) \cite{lxc}, become an alternative to traditional virtual machines because of their agility. The primary motivations for containers to be increasingly adopted are their conveniency to encapsulate, deploy, and isolate applications, lightweight operations, as well as efficiency and flexibility in resource sharing. Instead of installing the operating system as well as all the necessary softwares in a virtual machine, a docker images can be easily built with a Dockerfile \cite{docker}, which specifies the initial tasks when the docker image starts to run. Besides, container saves storage by allowing different running containers to share the same image. In other words, a new image can be created on top of an existing one by adding another layer. Compared to traditional virtual machines, containers provide more flexibility and versatility to improve resource utilization. Since the hardware resources, such as CPU and memory, will be returned to the operating sytem immediately. Because there is no virtualization layer in a container, it also incurs less performance overhead on the applications. Therefore, many new applications are programmed into containers. 

Performance comparison between containers and virtual machines has attracted many researchers \cite{desai2016survey, pahl2015containerization, soltesz2007container, babu2014system, felter2015updated, bhimani2017accelerating, seo2014performance, morabito2015power, xavier2014performance, sharma2016containers}. For example, researchers from IBM \cite{felter2015updated} have compared the performance of a docker container and a KVM virtual machine by running benchmarks such as Linpack, Stream, Fio, and Netperf. Their work was focusing on single container or virtual machine performance, and many of the workloads are micro benchmarks, which are very different from the big data workloads running in the cloud. Efforts have also been made to explore the era of using Docker containers in various cloud platforms, such as PaaS clouds and IoT clouds \cite{bernstein2014containers, celesti2016exploring}, but there lacks the performance study. Although some pioneers investigated the performance of containers and virtual machines in the big data era \cite{seo2014performance, bhimani2017accelerating},  they either lacks in-depth scalability analysis, which is a critical cloud performance criteria, or their experimental environment contains only one physical machine, which is not a representative setup of a cloud. Therefore, how the big data workloads perform and scale in a container cloud versus a virtual machine cloud remains to be an open issue.  

In this paper we conduct an extensive comparison study between containers and virtual machines in big data environment with three objectives: First, we investigate how much conveniency containers and virtual machines can bring to the system administrators. Specifically, we focus on how fast a big data computing environment can be setup from the scratch. A shorter setup time not only indicates an easier job for the system administrators, but also brings shorter latency, thus better experience to the users. This is an important factor that the cloud service providers usually take into their considerations. Second, we measure the impact of using containers and virtual machines on the performance and scalability of different big data workloads. Each big data job consists of multiple executors, and all the executors are distributed in the containers or virtual machines. The performance of the big data jobs studied in this paper can also reflect that of many other applications running inside a container or a virtual machine. Third, we analyze the reasons why containers and VMs can have different impacts on the big data workloads. In order to have a deeper understanding, we collected and compared many runtime system level metrics, such as resources utilization. We believe that this study will benefit both scientists and application developers to make more informed design decisions in terms of configuring, tuning, and scaling big data cloud.

%we plan to perform a comprehensive study on the Spark performance impact on different platforms (i.e. virtual machine and dockers container), this includes different kinds of jobs, workloads and different CPU and memory utilities on work nodes. Second, we plan to provide experimental evaluation and analysis on the performance impact of tuning configuration parameters on application performance. As a result of this study, we attempt to answer a number of challenging and yet most frequently asked questions regarding big data applications on Dockers container through extensive and in-depth experimentation and measurements. Example questions include but not limited to: Why and how exactly Dockers container have better configuration and deployment performance than virtual machine?  What kinds of Spark jobs are more suitable for Dockers containers to carry out a high performance distributed results?  Can a Spark cluster on Dockers container works effectively when datasets and data analysis jobs demand larger CPU and memory resources than the available allocations?  To the best of our knowledge, this is the first in-depth measurement study about Spark performance tuning of data analytic services between virtual machine and dockers container.  We conjecture that this study will benefit both scientists and application developers to make more informed design decisions in terms of configurations, performance optimization parameters, and domain specific tuning in terms of specific operations in scaling large scale data analytics applications.

The rest of the paper is organized as following: the background of the container technology and its comparison to the virtual machine technology, as well as the Spark big data platform, are introduced in section \ref{background}, section \ref{relatedwork} discusses the related researches efforts, the experimental setup and results are shown and analyzed in section \ref{analysis}, and the paper is concluded by section \ref{conclusion}.

\section{Background} \label{background}
In this section, we first introduce the principle of the container technology, compare its architecture with that of the virtual machines, and explains why it is becoming increasingly popular. Then, we briefly explain Spark \cite{zaharia2010spark}, which is used as a big data platform in this study.

\subsection{Container v.s. Virtual Machine(VM)}
A container is similar to an application, which runs as a process on top of the operating system(OS) and isolates with each other by running in its own address space. Nevertheless, more than a normal process, a container not only includes the application executable, but also packs together all the necessary softwares that the application needs to run with, such as the libraries and the other dependencies. \emph{Cgroups}  \cite{cgroups} and \emph{namespaces} \cite{namespaces} are two particular features that enables different containers running on the same physical machine. \emph{Cgroups}, also named as control groups, is a Linux kernel mechanism to control the resources allocation. It allows system administrators to allocate resources such as CPU, memory, network, or any combination of them, to the running containers. The resources allocated to each container can be adjusted dynamically, and the container cannot use more resources than being specified in the \emph{cgroups}. \emph{Namespace} provides an abstraction of the kernel resources, such as process IDs, network interfaces, host names, so that each container appears to have is own isolated resources. With \emph{namespaces}, for instance, processes running in different containers will not conflict with each other even when they have the same process ID.

\begin{figure}[htb]
\centering
\includegraphics[width=2.5in, height=1.8in]{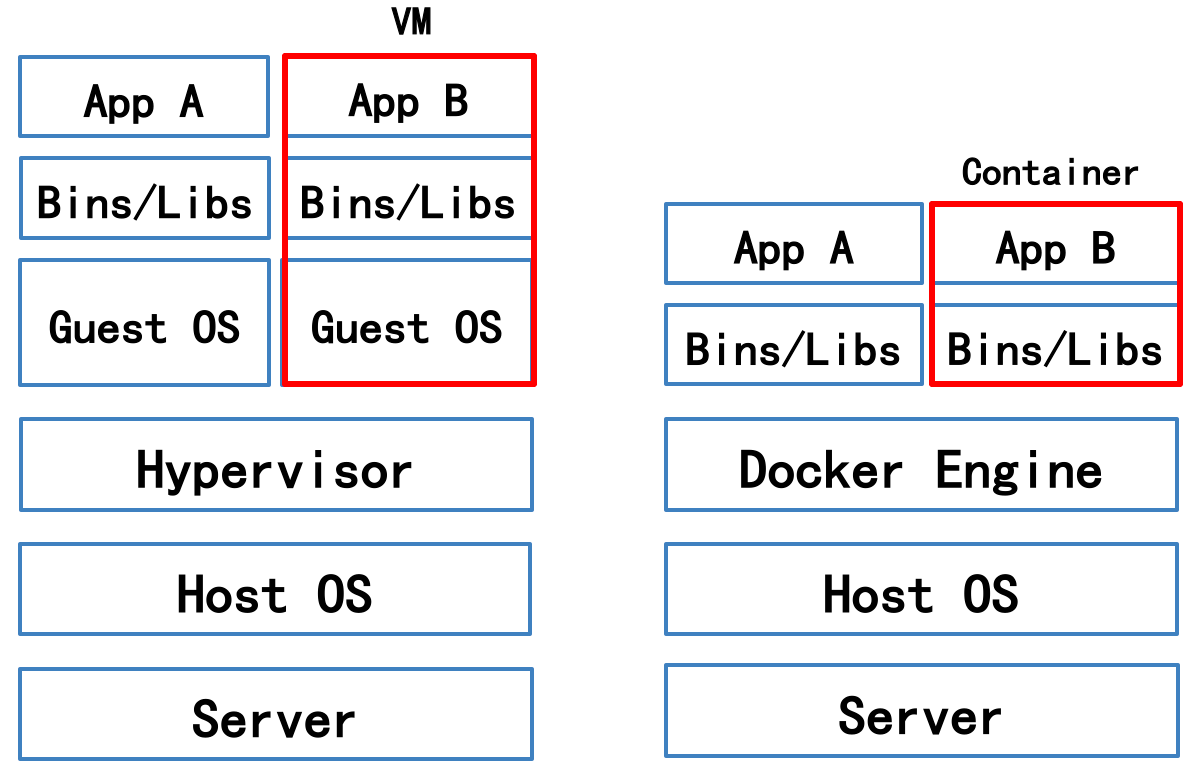}
\caption{\label{kvm_docker} Architecture comparison virtual machine v.s. container}
\end{figure}

Figure \ref{kvm_docker} shows the architecture comparison between a container and a VM. There are three major differences between these two technologies. First, a container is more lightweight than a VM. A container includes only the executables and its dependencies and different containers on the same machine share the same OS(operating system). While a VM contains a full set of OS, and different VMs do not share the OS. A VM can run an OS that is different from its host machine, while the container needs to use the same OS as the host. Second, the hypervisor, such as VMware ESXi \cite{vmware} and KVM \cite{kivity2007kvm}, is necessary in a VM environment, while it is not required for containers. For a VM, it needs to act as an independent machine which has the control of all its resources. However, in fact, a VM runs in a non-priviledged mode which does not have the capability to execute many privileged instructions. Therefore, a hypervisor is needed to translate a VM instruction into an instruction that can be executed by the host. On the contrary, since a container does not need to execute any priviledged instruction, it communicates with the OS through the system calls, thus  no other layer is required in between. Third, each VM has its own image file, while different containers may share some of their images. More specifically, container images are created in a layered manner, in which a new image can be created upon an existing image by adding another layer that contains the difference between the two. The image files of different VMs are isolated from each other.

Researchers and practitioners are paying increasing interest and attentions to the container technology for a number of reasons. On one hand, using container is more cost effective. As a lightweight solution, the size of a container is usually within tens of MB while that of a VM can take several GB. Also, to run the same application, a container usually takes less hardware resources since it does not need maintain an OS. On the other hand, since there is no hypervisor, containers are able to provide better application performance, especially when the applications need to talk to the I/O devices.

\subsection{Spark}
Apache Spark \cite{zaharia2010spark} is an open source cluster computing system. In contrast to Hadoop's two-stage disk-based MapReduce paradigm, Spark provides a resilient distributed data set (RDD) and caches the data sets in memory across cluster nodes. Therefore, it can efficiently support a variety of compute-intensive tasks, including interactive queries, streaming, machine learning, and graph processing. Spark is introduced as a unified programming model and engine for big data applications. It is gaining popularity as a big data cloud platform.

\section{Related Work}\label{relatedwork}
With the increasing popularity of container technologies, many research efforts have been made to explore the advantages of containers, improve container performance and security, as well as to compare the containers with the VMs.

Carl \cite{boettiger2015introduction} explored the common challenges that prevent the reproducibility of many research projects, and examined how the Docker container technology can solve them. For example, many projects require specific dependencies to reproduce similar results as the original researchers, but it is difficult to simply provide a installation script due to different underlying OS and hardwares the project is running on. Docker solves this issue by providing a lightweight binary image in which all the softwares have already been installed. Paolo et al. \cite{di2015impact} investigated the advantages of using Docker to chain together the execution of multiple containers to run the pipeline application such as Genomic pipelines. Douglas et al. \cite{jacobsen2015contain} suggested that the flexibility and the reproducibility of the containers will drive its adoption in HPC environments. Many researches have also discussed the advantages of using the container technology in cloud computing. For example, David \cite{bernstein2014containers} proposed that the Platform-As-A-Service(PaaS) cloud can benefit from containers by its easy deployment, configuration, as well as convenient management of the applications in the cloud.

In terms of container performance and security, Roberto \cite{morabito2016performance} presented a performance evaluation of using containers in the field of Internet of Things. By running containers on top of a single board computer device such as Raspberry Pi 2, Roberto showed an almost negligible impact of the containerized layer compared to native execution. Miguel et al. \cite{xavier2013performance} revealed the possibility of using container technologies, such as Linux VServer, OpenVZ, and Linux Containers, to achieve a very low overhead HPC environment compared to the native setups. Thanh \cite{bui2015analysis} analyzed the security of Docker from its internal mechanisms as well as its interacts with the security features of the Linux kernel. Sergei et al. \cite{arnautov2016scone} enhanced the container security by using Intel SGX trusted execution support to prevent outside attacks.

Comparative study between containers and VMs has attracted many attentions \cite{desai2016survey, pahl2015containerization, soltesz2007container, babu2014system, felter2015updated, bhimani2017accelerating, seo2014performance, morabito2015power, xavier2014performance}. For example, Janki et al. \cite{bhimani2017accelerating} compares the performance of Spark jobs running in a container cluster and a VM cluster, but the cluster used in their experiments is fix-sized, and only one physical machine is used, which is not a representative setup of a cloud environment. Claus \cite{pahl2015containerization} compared the virtualization principle between containers and VMs. The author also illustrated containers as a more suitable technology to package applications and manage the PaaS cloud, due to the better support of microservices from the containers. Wes et al. \cite{felter2015updated} used a suite of benchmark workloads, such as Linpackto, Stream, Fio, and Netperf, to compare the performance of containers and VMs by stressing their CPU, memory, storage, and networking respectively. Their results showed that both VMs and containers introduced negligible overhead for CPU and memory performance, while container performs better for disk and network I/O intensive applications, especially when there is random disk I/O. Kyoung-Taek et al. \cite{seo2014performance} briefly compared container and VM boot up time, as well as their computational performance by letting each of them compute 100000 factorial. Roberto \cite{morabito2015power} did a comparative study of power consumption between containers and VMs, and showed that both technologies consumed similar power in idle state and in CPU/Memory stress tests, while containers introduced lower power consumption for network intensive applications.

\section{Performance Study and Analysis} \label{analysis}

\subsection{Setup} \label{exp_setup}
Our experimental environment consists of a five-node cluster, and each node is a physical machine with 8 Intel i7-3770 3.40GHz CPU, 16GB memory, and 1TB disk. The Ubuntu 16.04 LTS 64-bit is running as the OS on the five machines as well as the in all the VMs. Spark 2.0.2 is used as the big data platform. Docker 1.12.6 and KVM 2.5.0 are used to manage the containers and VMs. As shown in the Figure \ref{exp_setup}, each of the five physical machine is equipped with a Realtek RTL8111 network card, which are connected through a TP-LINK TL-SG1024DT network switch. Since Spark follows a master-slave architecture, we use one physical machine as a master node, while the other four machines as slave machines. All the Spark tasks are running in either containers or VMs. In the VM setup, a network bridge is created on each physical machine, which not only enables multiple VMs on the same physical machine to share the network card, but also allows VMs on different physical machines to communicate with each other. While Docker containers are running, Flannel \cite{flannel} is used for containers to communicate across physical machines. Flannel runs as an agent on each machine and allocates a subnet lease out of a larger and preconfigured network address space. As also shown in Figure \ref{exp_setup}, there is a \emph{Collectd} \cite{collectd} daemon running on each of the slave machine, and a InfluxDB \cite{influxdb} running the master machine. The daemon collects various system metrics, such as CPU and memory utilization, every single second, and writes them as the time series data into the InfluxDB.

\begin{figure}[htb]
\centering
\includegraphics[width=2.8in, height=2.2in]{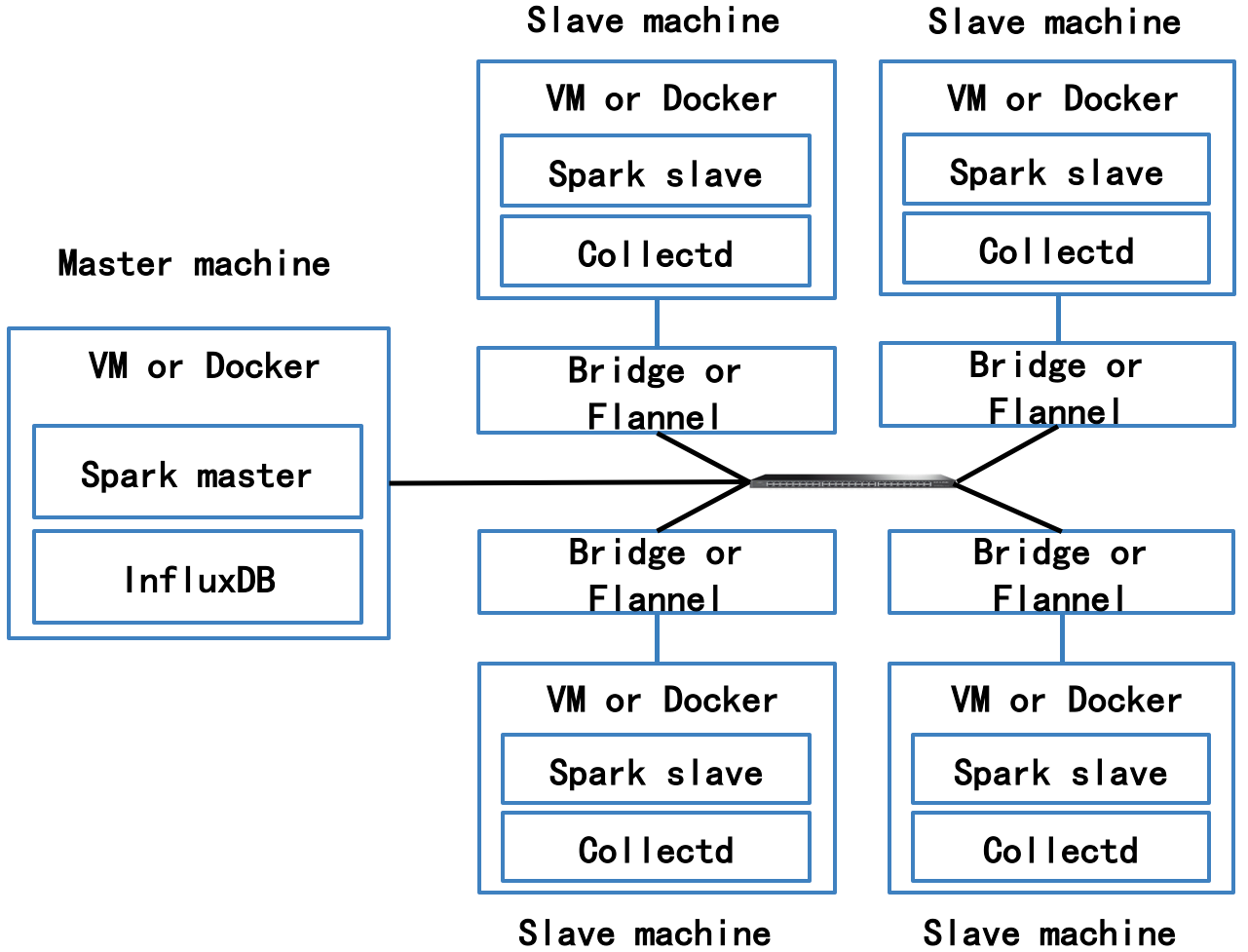}
\caption{\label{exp_setup} Experimental setup}
\end{figure}

The workloads we used in this paper and their configurations are listed as follows:
\begin{itemize}
  %\item Bayes, a popular classification algorithm for knowledge discovery and data mining, with 1 million pages and 100 classes. 
  \item Kmeans, a method of vector quantization for cluster analysis in data mining. The input data contains 20 dimensions, 5 million samples. The job runs for 5 iterations.
  \item Logistic Regression, a statistical method for analyzing a dataset in which there are one or more independent variables that determine an outcome. The input data includes 10 thousands examples and 20 thousands features.
  \item Pagerank, an algorithm that counts the number and quality of links to a page to determine a rough estimate of how important the website is. The input data includes 300 thousands pages, and the job runs for 3 iterations.
  \item SQL Join, a task that consumes two different data sets and join them together to find data pairs in both sets that share a same value. The input data has information about 90 million user visits and 1 million pages.
\end{itemize}

%For each workload, all the above mentioned resource utilization on each of the physical machine is recorded every second, and the average resource utilization during the whole execution is calculated. Figure \ref{} compares the average utilization of each resource among all the selected workloads.

\subsection{Deployment Conveniency}
Easy deployment is one of the benefits brought by virtualization technologies. For example, instead of doing all the setups from the scratch, a VM can be easily moved from one physical machine to another to create an identical setup for either backup or migration purposes. As a lightweight virtualization technology, how much convenience can containers bring to system administrators is the question that worth being investigated.

In this subsection, we compare the time spent on creating a virtualized and containerized Spark cluster on a single physical machine. Each cluster consists of 3 VMs or 3 Docker containers respectively. The cluster setup is divided into three steps in both cases. The first step is building images. In the VM case, virt-manager \cite{virtmanager} is used as a graphic tool to create a VM image from an ISO file. This step includes creating the disk image file and installing the VM OS. Since each VM runs on its own image, the image file needs to be copied 3 times in order to setup a three-VM cluster. In the container cluster case, Dockerfile \cite{docker} is used to create docker images. According to the Dockerfile, the Docker engine automatically downloads an Ubuntu base image from the Docker repository, and then executes the commands specified in the Dockerfile to copy the necessary files to the specific locations in the image. After being created, the size of each VM image is 4.1GB, while that of a single container image is 1.1GB, which is only 1/4 of that of a VM image. Besides, since Docker image is organized in a shared and layered manner, a new Docker image can be created based on an existing image, and only the incremental bytes need to be saved to the disk. The second step is setting up the Spark environment in each running instance. This step includes installing the JDK(Java Development Kit), configuring the Spark runtime, enabling SSH communication without using password, and etc. This step is almost the same no matter in a VM or in a Docker container environment. The only difference is that in the VM, we create a script which will be automatically executed after the VM starts. Meanwhile, for a docker container, the software environment is setup on the host machine beforehand, so that the Dockerfile can clone such environment into the container image accordingly. The third step is starting the Spark cluster using the scripts shipped with the Spark source code. Since staring the Spark executable on each node is independent of whether the executable is running in a VM or a container, this step is exactly the same for the two scenarios.

In order to compare the efficiency between setting up a VM Spark cluster and a container Spark cluster, we put the above steps into documents, ask 15 students to setup both clusters by following the instructions in the documents. The total time as well as time spent on each step is recorded. The students we choose are from a BigData class for graduates, they are familiar with how Spark works and they have previous experiences on setting up a Spark cluster. The students are also required to read through the documents beforehand to make sure they understand every step. Therefore, the time they spent can mostly reflect the complexity of setting up a VM cluster and a container cluster.

As shown in table \ref{deployment}, it takes 23 minutes to create a cluster using Docker containers, while using VMs takes 46 minutes, which is 100\% longer. By investigating the time taken by each of the three steps, we observe that containers save most of its time during the image building phase. There are two reasons. First, since containers share the same OS with the physical machine, installing the OS can be skipped. Second, since image files can be shared among different containers, only one image file needs to be built to setup the cluster. While using VMs, however, a image file needs to be copied multiple times, since each VM needs its own image to start with.The time spent on setting up Spark and starting cluster is similar between using containers and VMs.

\begin{table}
\centering
\caption{Comparison of time spent on building a three-node Spark cluster using containers v.s. virtual machines (miniutes)}
\label{deployment}
\begin{tabular}{lllll}
\hline
\multicolumn{1}{|l|}{}          & \multicolumn{1}{l|}{Total} & \multicolumn{1}{l|}{Build Image} & \multicolumn{1}{l|}{Setup Spark} & \multicolumn{1}{l|}{Start Cluster} \\ \hline
\multicolumn{1}{|l|}{VM}        & \multicolumn{1}{l|}{46}      & \multicolumn{1}{l|}{28}            & \multicolumn{1}{l|}{13}            & \multicolumn{1}{l|}{5}              \\ \hline
\multicolumn{1}{|l|}{Container} & \multicolumn{1}{l|}{23}      & \multicolumn{1}{l|}{6}            & \multicolumn{1}{l|}{12}            & \multicolumn{1}{l|}{5}              \\ \hline
                                &                            &                                  &                                  &                                   
\end{tabular}
\vspace {-0.5cm}
\end{table}

\subsection{Bootup Efficiency} \label{bootup}
Bootup latency refers to the period from starting a VM or container to the time that it can provide services to clients. This is an important factor that cloud service providers usually consider. There are several reasons. On one hand, a shorter bootup latency means a smaller latency to provide services, which brings better user experiences. On the other hand, faster bootup also allows more elastic resource allocation, since the resources do not have to be reserved for the VM or Docker that is starting up. Therefore, in this subsection, we show the difference of the bootup latency between Docker containers and VMs, and also investigate the reasons that lead to such difference.

\begin{figure}[htb]
\centering
\includegraphics[width=2.5in, height=1.8in]{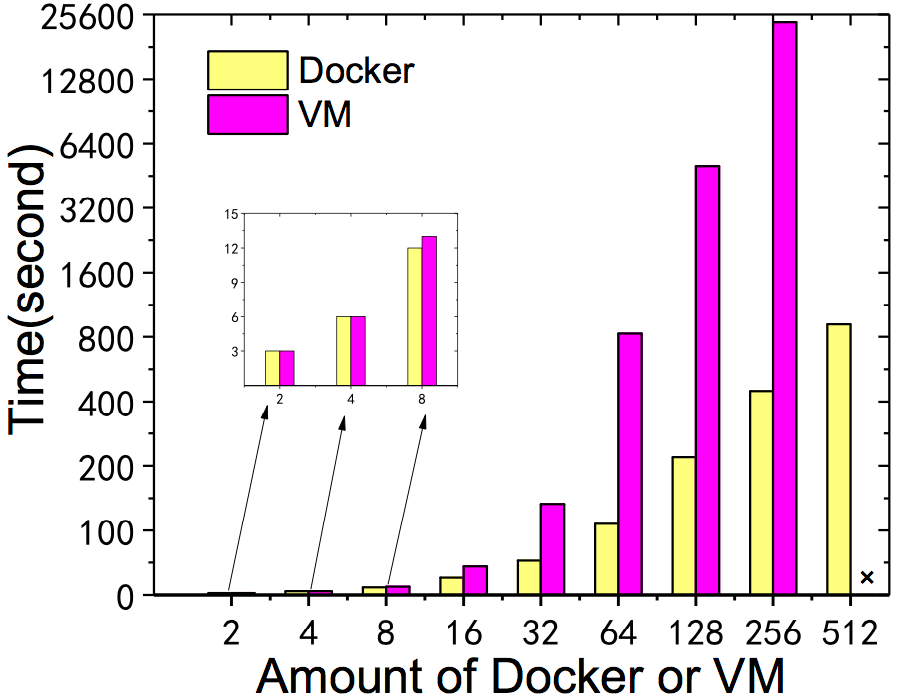}
\caption{\label{boot_up_latency} Bootup latency (cross in the figure means failed to bootup that amount of virtual machines).}
\end{figure}

Figure \ref{boot_up_latency} compares the bootup latency of Docker containers and VMs by varying the number of instances that boot up. In each set of experiments, all the instances are started in a sequential manner, and we measure the time between the first instance starts up and the last instance finishes booting up. Note that no applications is initialized after the boot up. There are several interesting observations: on one hand, the physical machine can hold more Docker containers than VMs. We noticed that it is very difficult to start new VMs when there are around 250 idle VMs on the host. For example, it takes more than 1000 seconds to boot up one VM, and the physical machine responses very slowly and is almost not useable. On the contrary, it takes only 987 seconds to start 512 Docker containers, which is on average 1.89 seconds per container bootup. On the other hand, to start the same amount of instances, Docker containers take much less time than the VMs, especially when the number of instances is large. We can see that when there are only 2 Docker containers or VMs, their bootup latency is almost identical. However, when this number increases to 256, Docker containers takes 479 seconds while the VM spends 24295 seconds, which is 50.72 times longer.

\begin{figure}[htb]
\centering
\subfigure[\small VM]{\label{vm_boot_memory} \includegraphics[width=1.6in, height=1.5in]{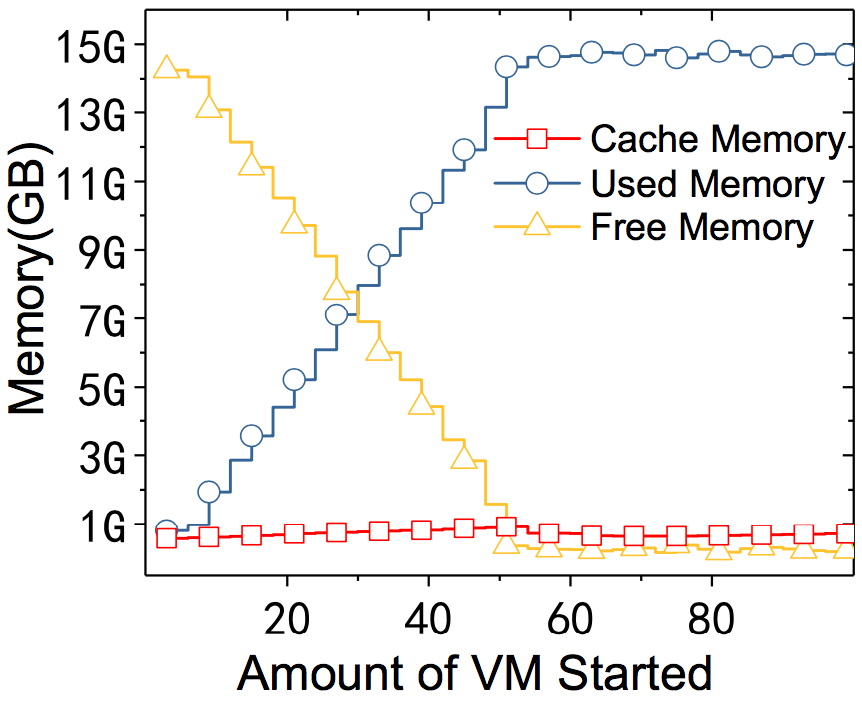}}
\subfigure[\small Docker container]{\label{docker_boot_memory} \includegraphics[width=1.6in, height=1.5in]{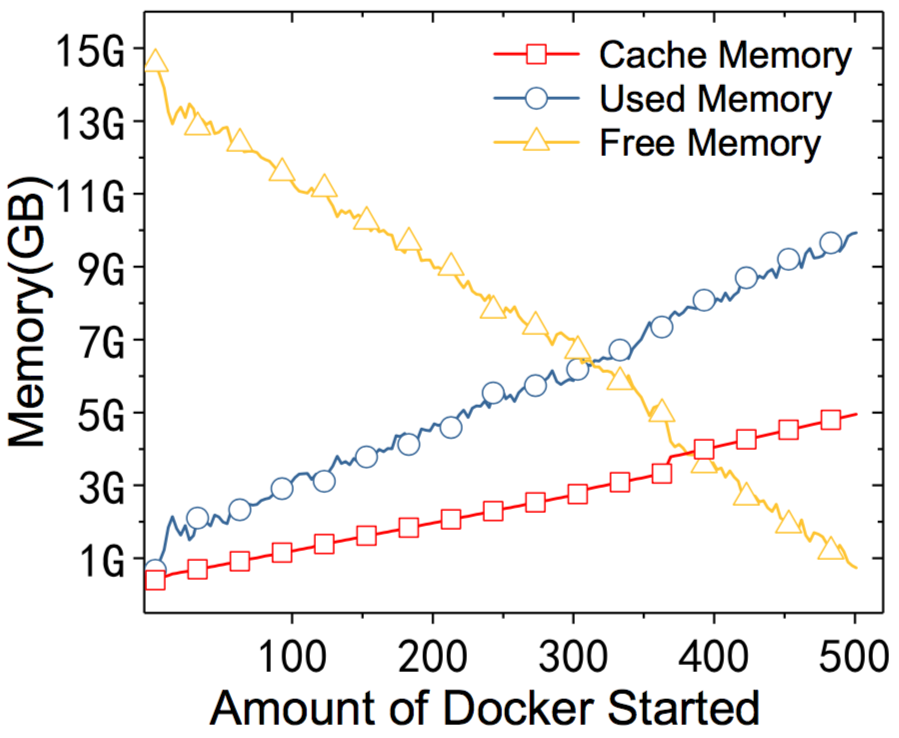}}
\caption{\label{boot_mem} Physical machine memory utilization during the virtual machine or docker container bootup.}
\end{figure}

In order to gain a better understanding of why there is such a huge difference between the Docker and VM bootup efficiency, we measured the memory usage of the physical machine, and the results are shown in Figure \ref{boot_mem}. We also collected the CPU and disk I/O statistics but decided not to show them here. This is because the CPU utilization is pretty low during the whole process, and there is not much disk I/O contention since all the instances are started sequentially. There are several interesting observations from this figure. First, both VM and Docker container start with less memory than being specified. Figure \ref{vm_boot_memory} shows, on average, a VM uses 0.23GB memory after it boots up, although each instance is allowed to take up to 2GB memory. This mechanism allocates resources to a VM or Docker container only when needed, which minimizes the idle resources in each instance. Second, a Docker container takes less memory than a VM after it boot up. while a VM takes 0.23GB, Figure \ref{docker_boot_memory} shows a Docker container only takes 0.03GB memory. This is because a VM needs to maintain an operating system while a container can share the same operating system with the physical machine. This is also the reason why the same physical machine can hold larger numbers of containers than VMs. 

%Third, memory cached is used more often in starting Docker containers than booting up VMs. By comparing Figure \ref{} and Figure \ref{}, we observe that the amount of used memory cache keeps consistent when VMs boot up, while that number increases linearly with the number of started containers. Such difference is due to the fact that the container image files are used in a shared manner and organized as layers \cite{} while each VM has its own image file. In this set of experiment, all the containers share the same image file. When a new container starts, it adds 

\subsection{Application Performance}
\begin{figure*}[htb]
\centering
\subfigure[\small Kmeans]{\label{runtime_kmeans} \includegraphics[width=1.6in, height=1.6in]{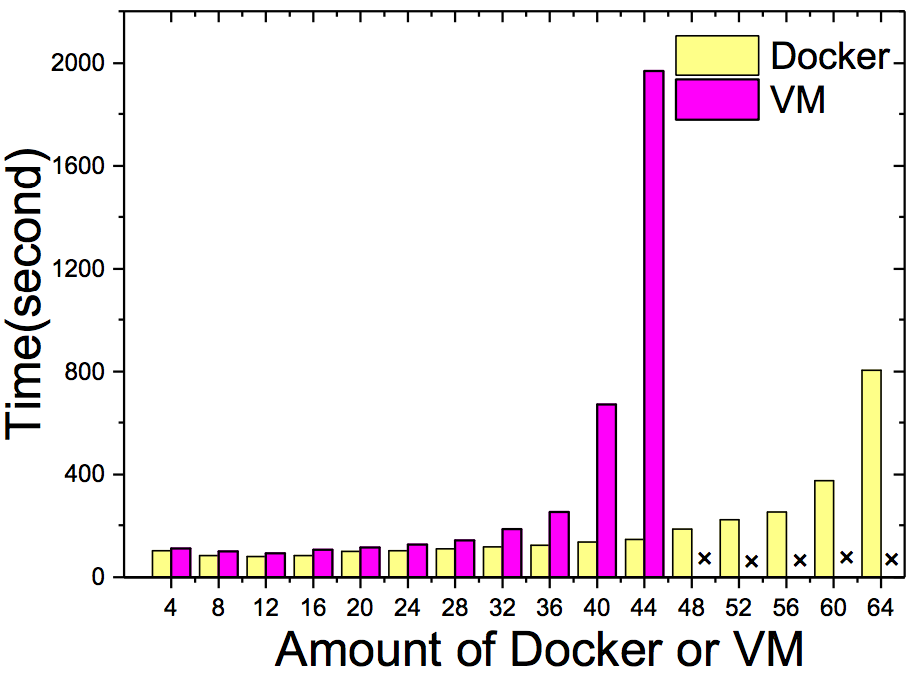}}
\subfigure[\small Logistic Regression]{\label{runtime_lr} \includegraphics[width=1.6in, height=1.6in]{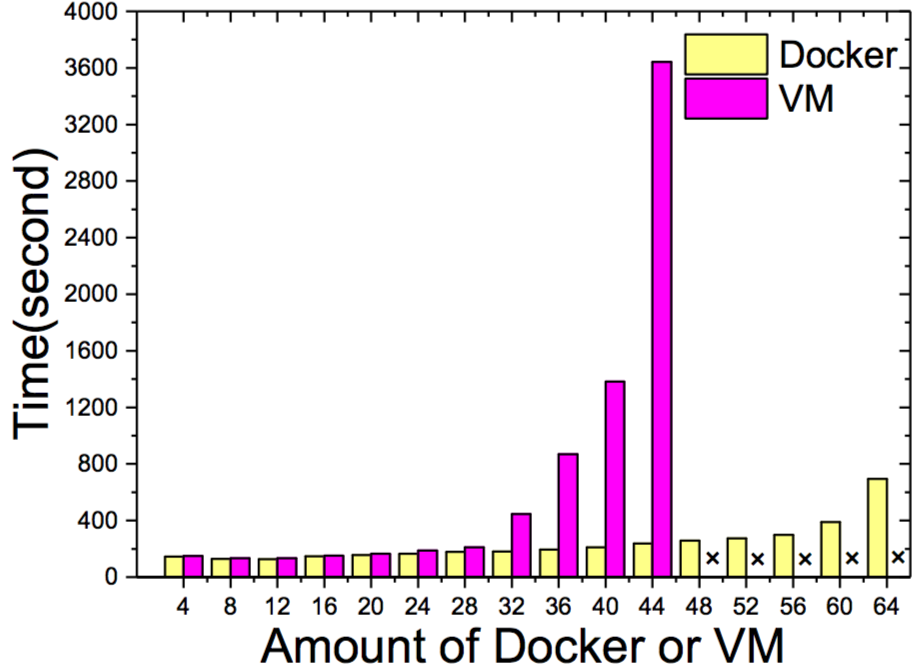}}
\subfigure[\small Pagerank]{\label{runtime_pg} \includegraphics[width=1.6in, height=1.6in]{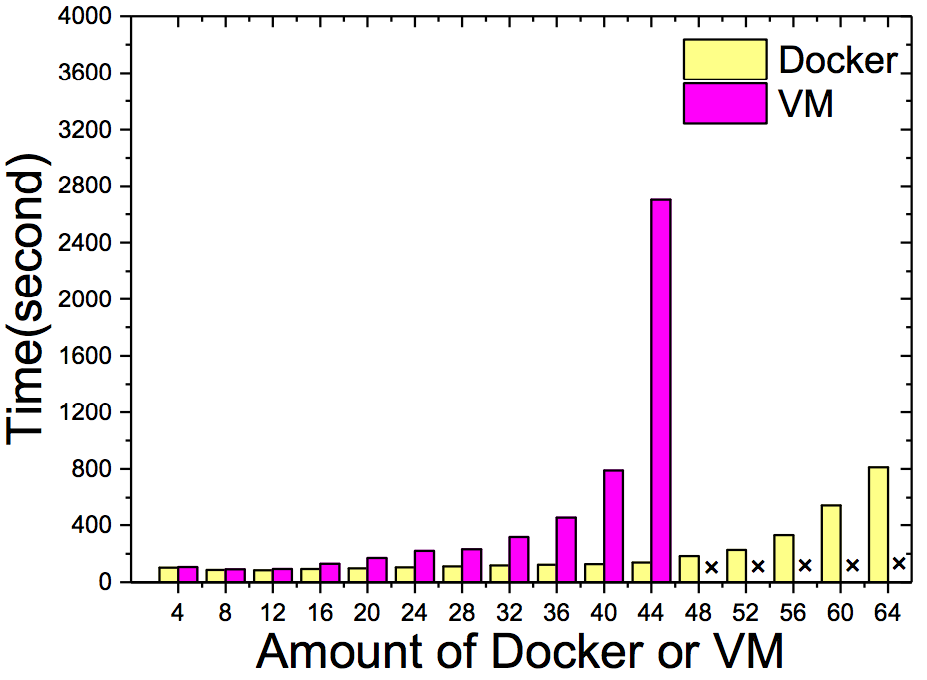}}
\subfigure[\small SQL Join]{\label{runtime_join} \includegraphics[width=1.6in, height=1.6in]{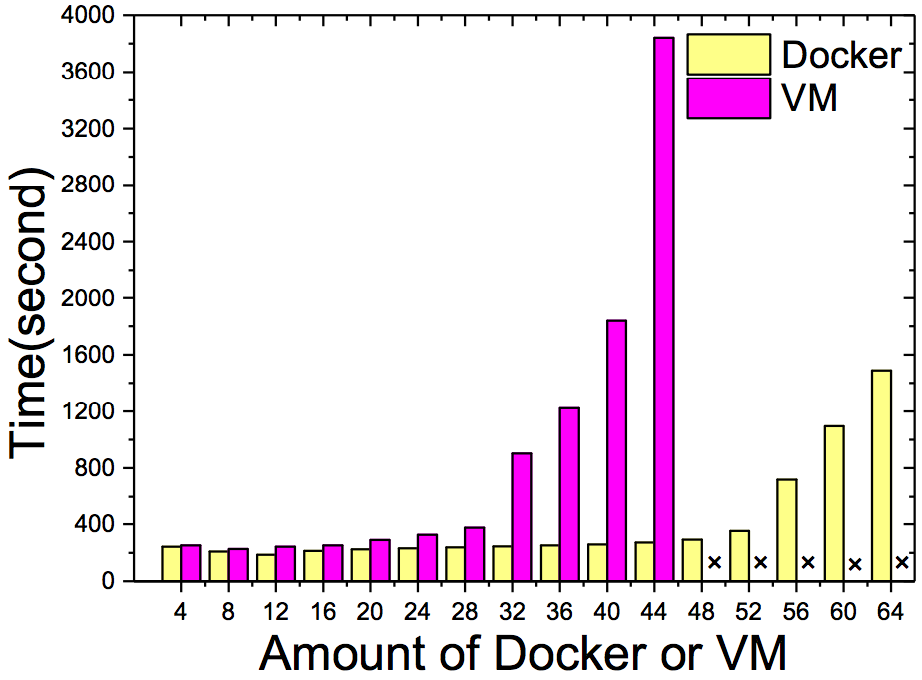}}
\caption{\label{runtime} Execution time of different Spark workloads running with containers and virtual machines (cross in the figure means the workload failed to finish).}
\end{figure*}

In this section, we measure and compare the performance and the scalability between a container environment and a VM environment. Different Spark workloads, as specified in section \ref{exp_setup}, are used as benchmarks. Specifically, in each set of the experiment, the size of the Spark cluster increases from 4 to 64 containers or VMs, and all the containers or VMs are equally distributed among all the slave machines. In other words, the number of containers or VMs running on the each slave machine changes from 1 to 16. Every single Spark workload runs with each size of the cluster, and the execution time is collected by an average of 5 runs. Figure \ref{runtime} shows the results, which compare the execution time of each workloads running in container environment and VM environment. 
\begin{figure*}[htb]
\centering
\subfigure[\small 2 containers/machine]{\label{8docker_cpu} \includegraphics[width=1.6in, height=1.6in]{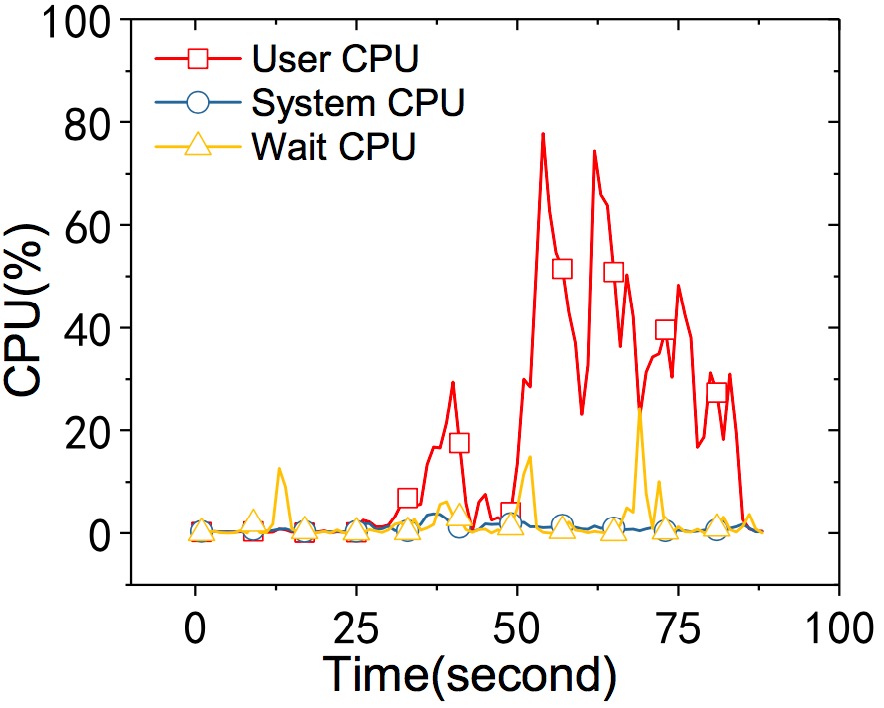}}
\subfigure[\small 4 containers/machine]{\label{16docker_cpu} \includegraphics[width=1.6in, height=1.6in]{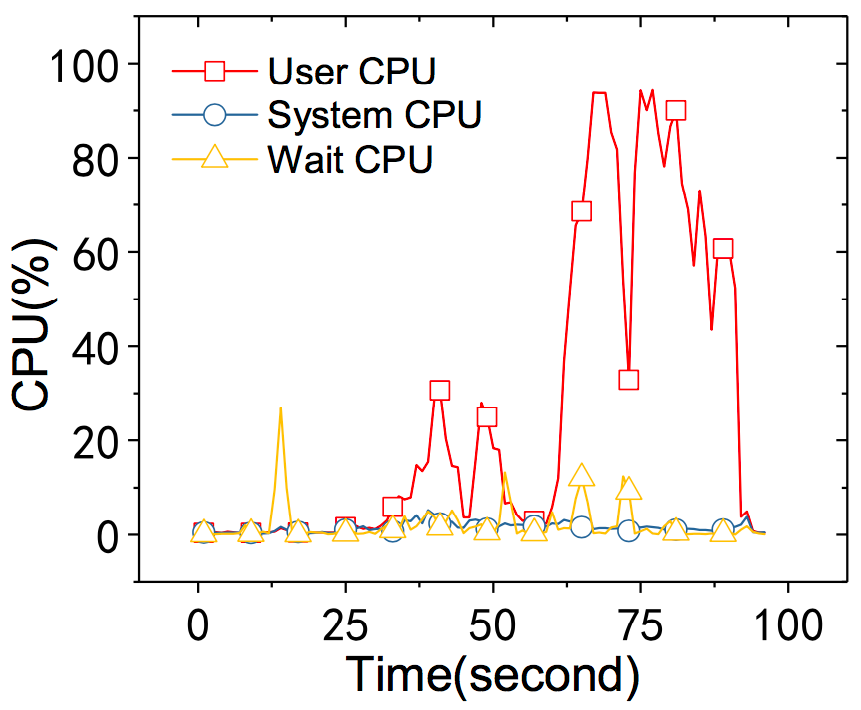}}
\subfigure[\small 8 containers/machine]{\label{32docker_cpu} \includegraphics[width=1.6in, height=1.6in]{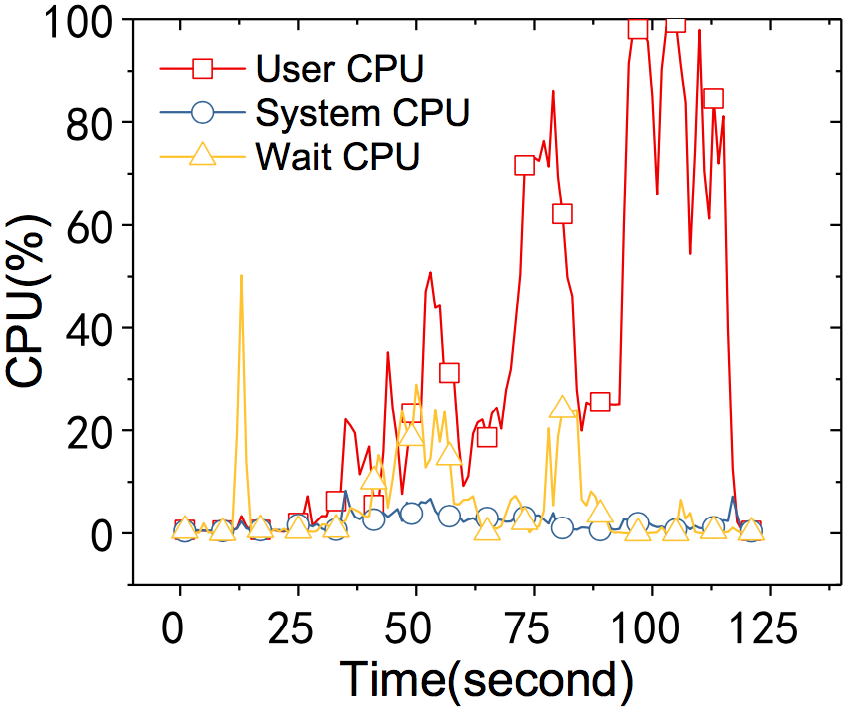}}
\subfigure[\small 12 containers/machine]{\label{48docker_cpu} \includegraphics[width=1.6in, height=1.6in]{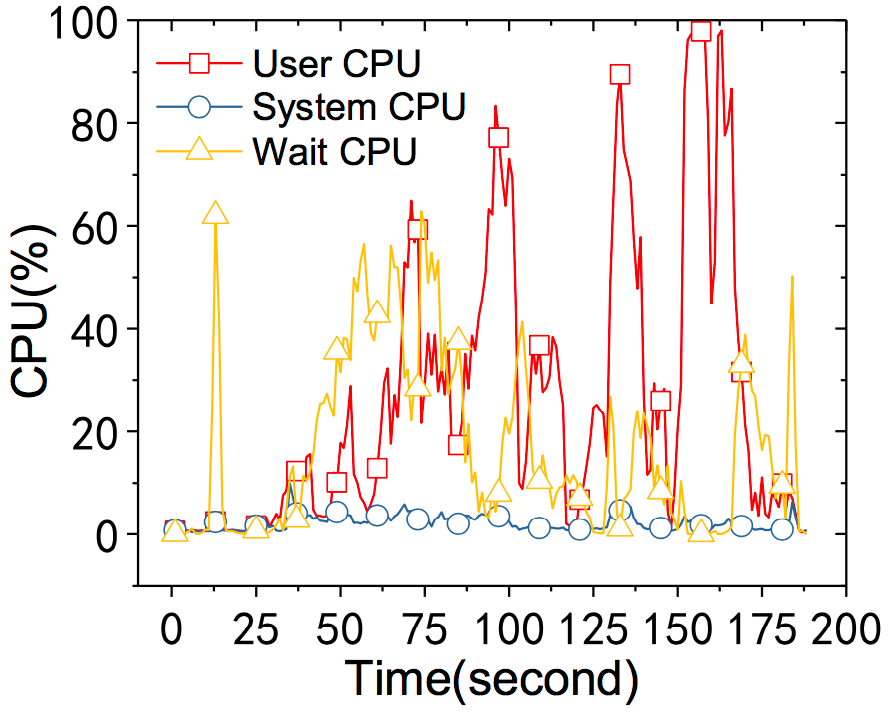}}
\subfigure[\small 2 VMs/machine]{\label{8vm_cpu} \includegraphics[width=1.6in, height=1.6in]{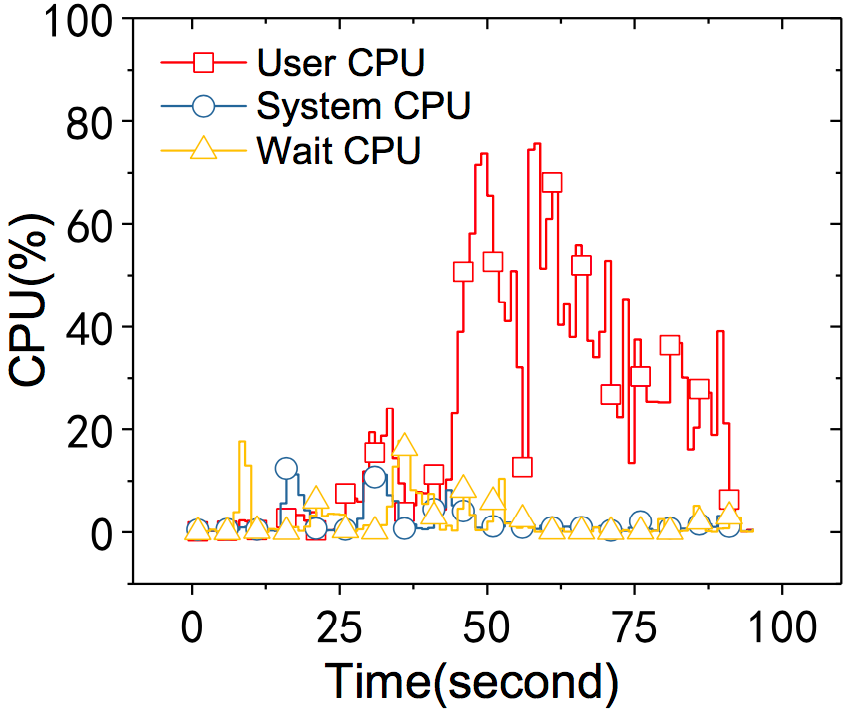}}
\subfigure[\small 4 VMs/machine]{\label{16vm_cpu} \includegraphics[width=1.6in, height=1.6in]{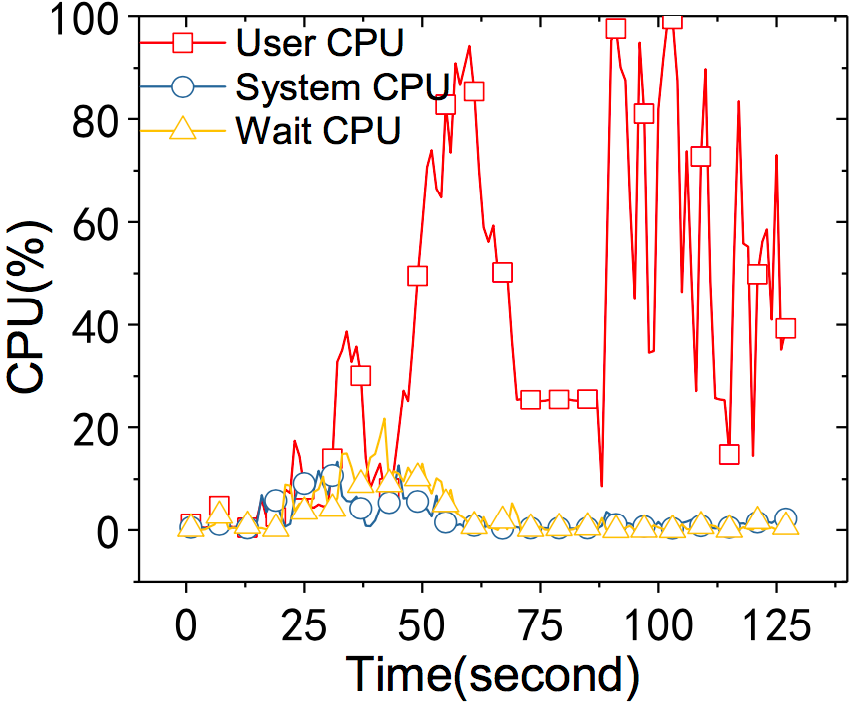}}
\subfigure[\small 8 VMs/machine]{\label{32vm_cpu} \includegraphics[width=1.6in, height=1.6in]{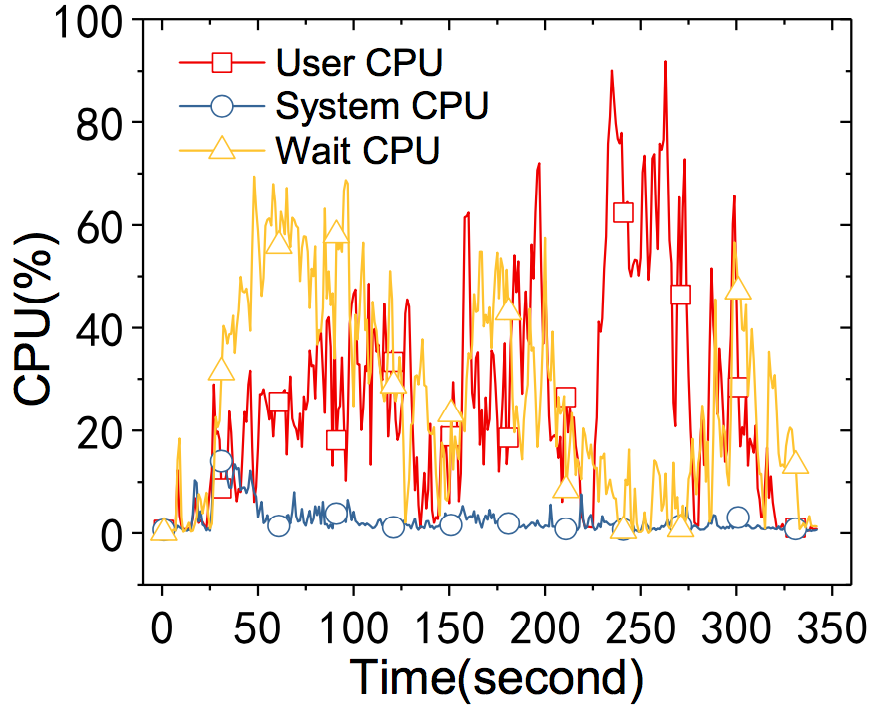}}
\subfigure[\small 12 VMs/machine]{\label{48vm_cpu} \includegraphics[width=1.6in, height=1.6in]{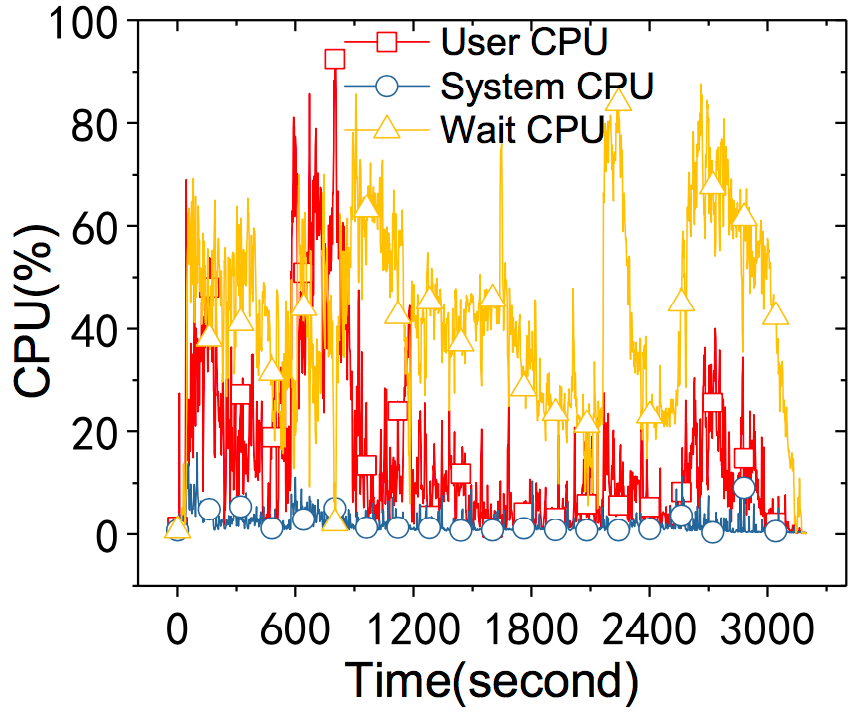}}
\caption{\label{cpu_comp} Average CPU utilization of all the slave machines running with containers and virtual machines, the workload is \emph{PageRank}.}
\end{figure*}
For every workload evaluated, its execution time decreases first and then increases with the growing number of the containers or VMs. Taking the \emph{SQL Join} workload for an example. Its execution time decreases from 242 seconds to 208 seconds when the total number of containers increases from 4 to 8. The execution time then further decreases to 186 seconds when the Spark size grows to 12 containers. Similar trend can also be observed when the VM is used. The decrease of the execution is because a Spark job usually consists of multiple tasks that can be parallelly executed in different containers or VMs. Given the same job with fixed number of tasks, the more containers or VMs to use, the higher parallelism the job has, which leads to shorter execution time. However, since each container or VM needs to access hardware resources, when the number of containers or VMs on the same physical machine increases, the resources contention becomes more intensive, which in turn brings impacts on the performance of the containers or VMs. This is why the execution time of \emph{SQL Join}, as well as the other three workloads shown in Figure \ref{runtime}, increases when more containers or VMs are added to the Spark cluster.
\begin{figure*}[htb]
\centering
\subfigure[\small 2 containers/machine]{\label{8docker_mem} \includegraphics[width=1.6in, height=1.6in]{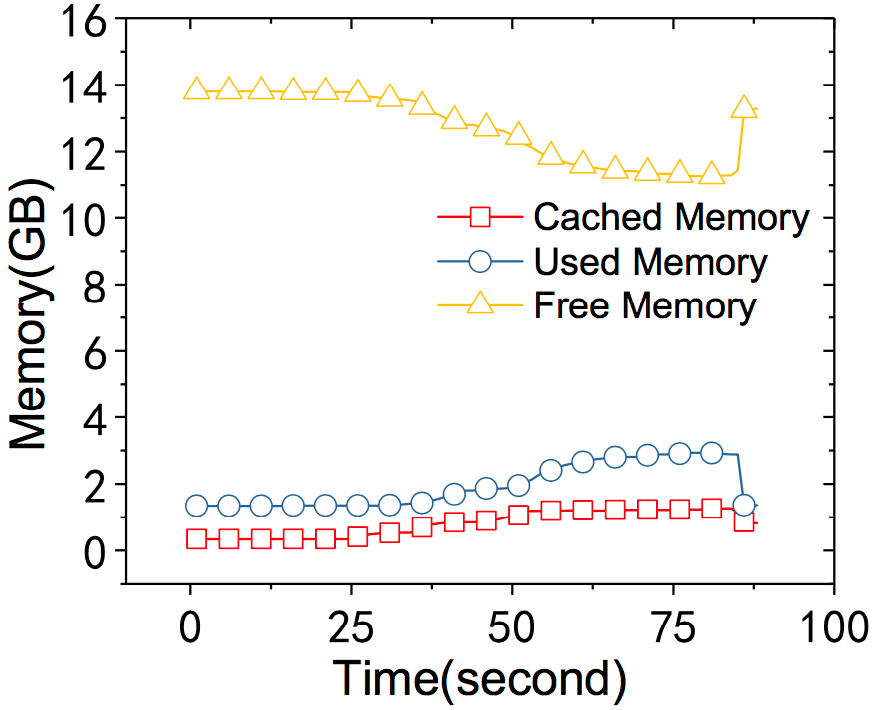}}
\subfigure[\small 4 containers/machine]{\label{16docker_mem} \includegraphics[width=1.6in, height=1.6in]{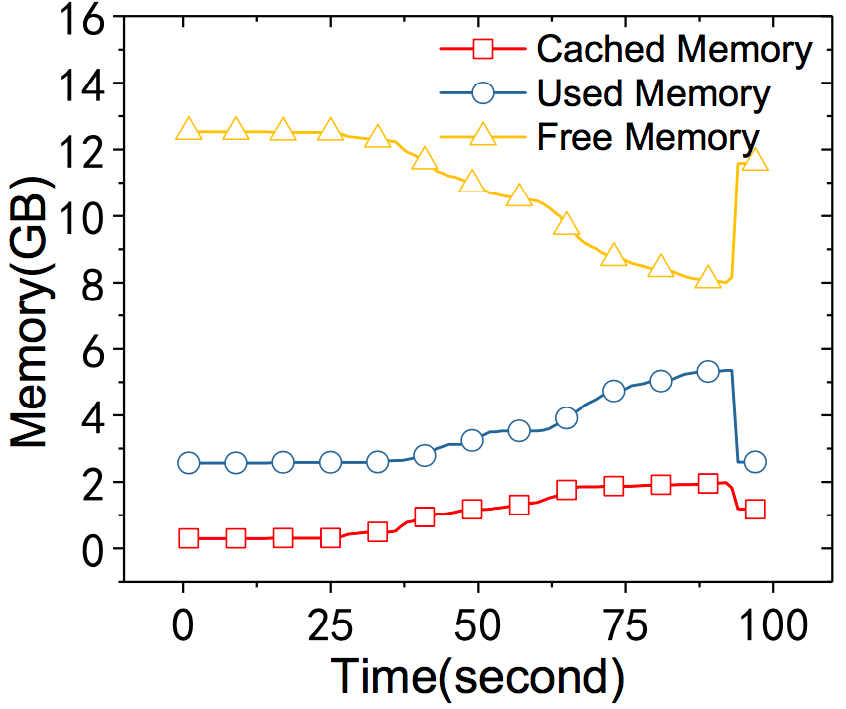}}
\subfigure[\small 8 containers/machine]{\label{32docker_mem} \includegraphics[width=1.6in, height=1.6in]{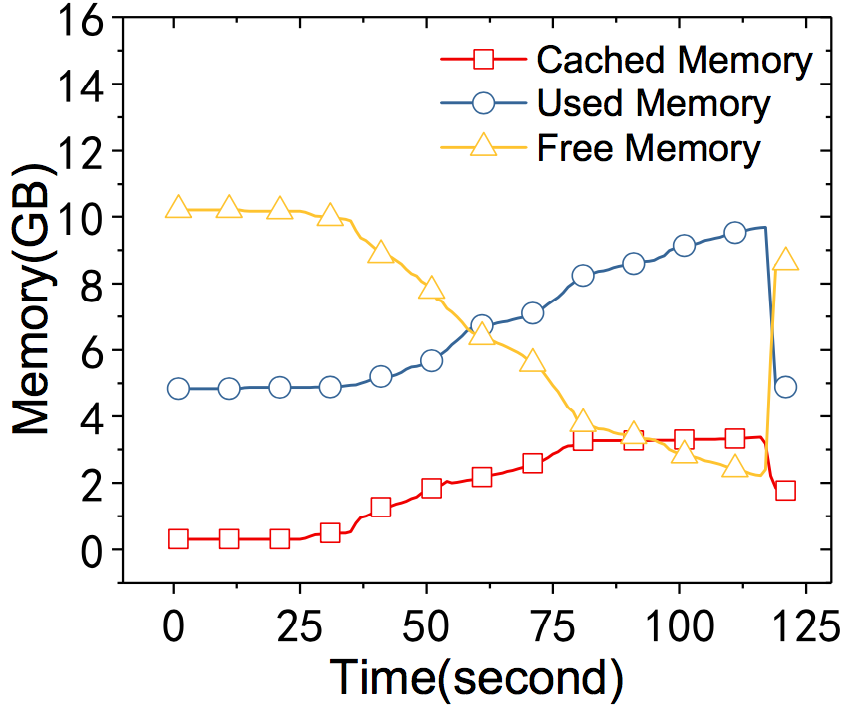}}
\subfigure[\small 12 containers/machine]{\label{48docker_mem} \includegraphics[width=1.6in, height=1.6in]{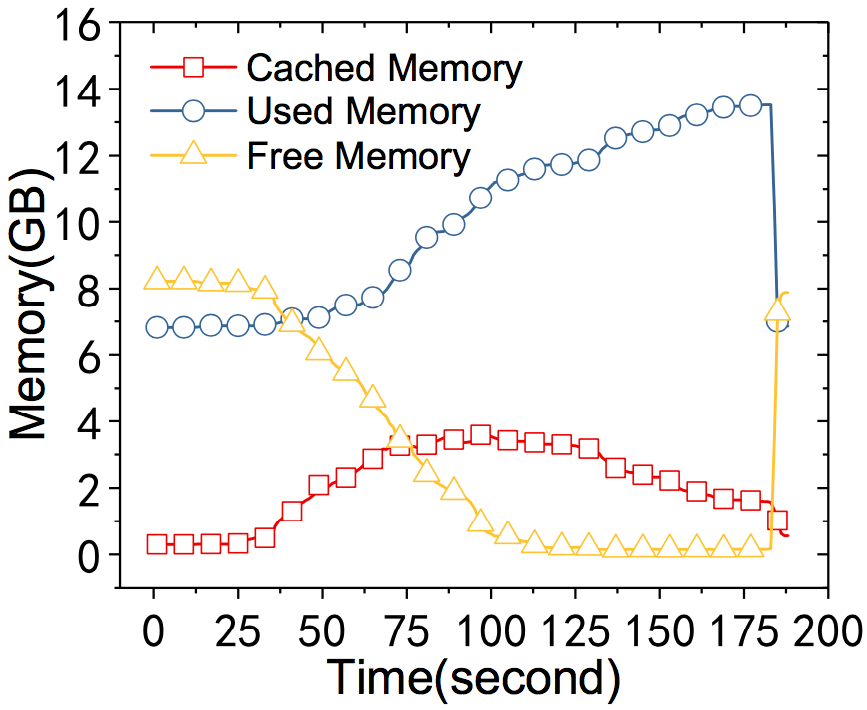}}
\subfigure[\small 2 VMs/machine]{\label{8vm_mem} \includegraphics[width=1.6in, height=1.6in]{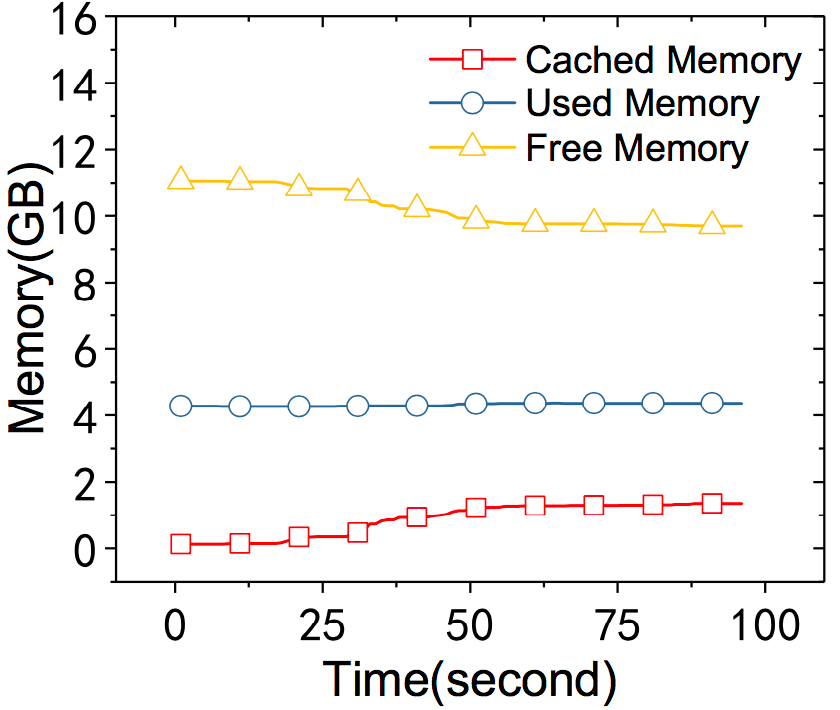}}
\subfigure[\small 4 VMs/machine]{\label{16vm_mem} \includegraphics[width=1.6in, height=1.6in]{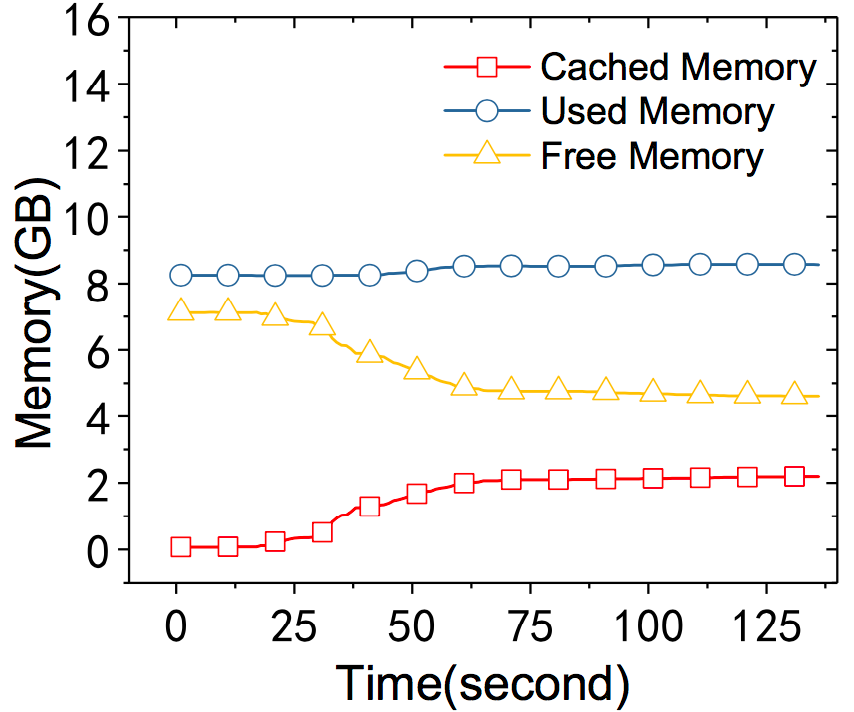}}
\subfigure[\small 8 VMs/machine]{\label{32vm_mem} \includegraphics[width=1.6in, height=1.6in]{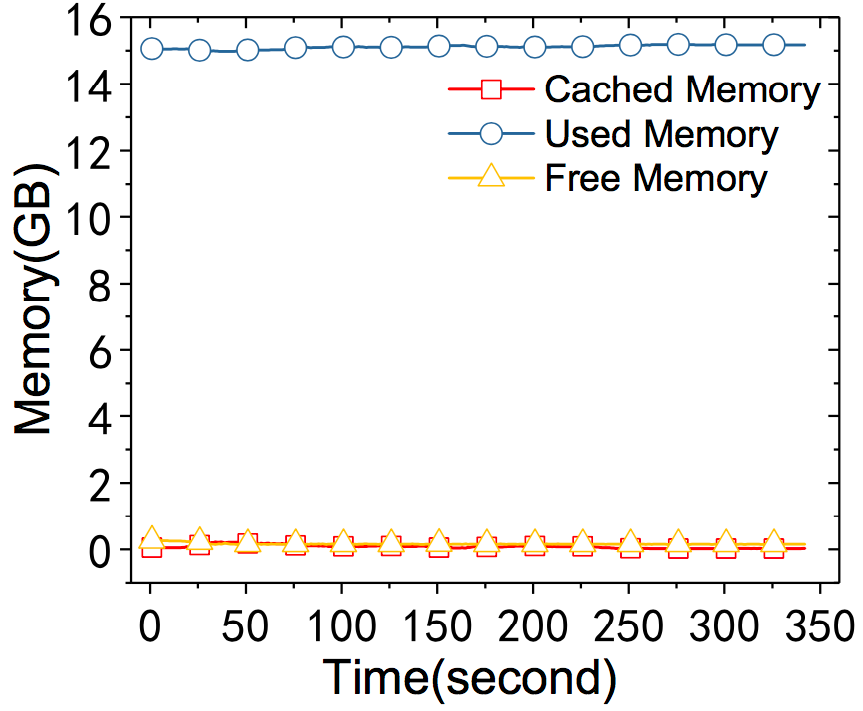}}
\subfigure[\small 12 VMs/machine]{\label{48vm_mem} \includegraphics[width=1.6in, height=1.6in]{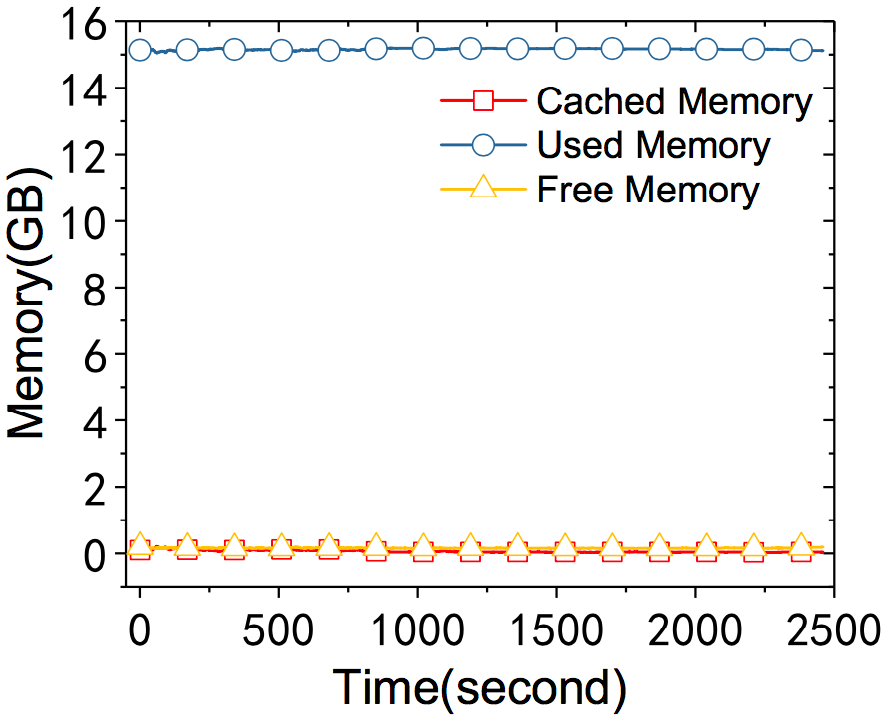}}
\caption{\label{mem_comp} Average memory utilization of all the slave machines running with containers and virtual machines, the workload is \emph{PageRank}.}
\end{figure*}
There is also an obvious difference in the execution time of the same workload between running with containers and VMs, which distinguishes the scalability between the two environments. First, for each single workload, its execution time is shorter in a container environment than that in a VM environment with the same size. For example, when \emph{PageRank} runs in a cluster size of 32, it takes 118 seconds to finish when containers are used while 316 seconds when the VMs are used. This result indicates that as a lightweight virtualization approach, containers bring less runtime overhead than the VM to the applications running inside. The reason is because both the hypervisor layer and the VM OS create additional overhead for the applications, in terms of context switch, privileged instruction translation, prolonged network or disk I/O path, and etc. Second, with the increase of the cluster size, containers provide a more scalable environment than VM. For \emph{Logistic Regression}, its execution time is 145 seconds in the container environment while 149 seconds in the VM environment when the cluster size is 4, the execution time only differs by 2.76\%. However, when the cluster size expands to 44, \emph{Logistic Regression} needs 238 seconds to finish in the container environment, whereas in the VM environment, it takes 3643 seconds, which is 14.31 times higher. Furthermore, when the cluster size increases to 48 or larger, the workload fails its execution due to too many aborted tasks in the VM environment. While in container environment, it can still successfully finish in a reasonable time.

In order to understand the reason why containers have a much better scalability than VMs, we collect system level metrics such as CPU and memory utilization on each of the slave machine, and the average number is shown in Figure \ref{cpu_comp} and \ref{mem_comp}. Figure \ref{cpu_comp} shows the average CPU utilization of the 4 slave machines with different cluster sizes while the \emph{PageRank} is running. The total CPU utilization is divided into three categories: user level CPU, system level CPU, and the wait time. We observe that the CPU is mostly spent on the user level and CPU wait, no matter in the container environment or in the VM environment, and no matter what the cluster size is. This is because the \emph{PageRank} workload is a user level application, and most of its instructions can run without being trapped into the kernel. A more important observation is that the CPU spent more time on wait with the increase of the cluster size, especially when VM is used. For instance, the average CPU wait time is less than 10\% in both container and VM environment when the cluster size is 8. However, when the cluster size increases to 48, the average CPU wait time in container environment is far less than that in the VM environment, which can be demonstrated by comparing Figure \ref{48docker_cpu} with Figure \ref{48vm_cpu}. In the VM environment, the user level CPU that can be utilized by each task is much less than that in the container environment. The high CPU wait time also indicates large amount of disk I/O with long latencies. Therefore, many tasks aborted due to timeout. Recall Figure \ref{runtime_pg}, this also explains why the \emph{PageRank} failed in a 48 VMs cluster while successfully finished in 183 seconds in a 48 containers cluster. 

The average memory utilization statistics of the 4 slave machines during the \emph{PageRank} execution is also collected and shown in Figure \ref{mem_comp}. The memory utilization is also divided into three categories: cached memory, used memory, and free memory. There are several interesting observations. First, comparing Figure \ref{8docker_mem} with Figure \ref{8vm_mem}, although 2GB memory is assigned to each container and VM to start with, 2 containers on each slave machine takes less than 2GB memory while 2 VMs takes about 4GB memory before the workload starts running. Note that different from the containers or VMs in section \ref{bootup}, in this set of experiments, they need to run a data generation job before the execution of the evaluated big data workloads This is largely because a container does not need to main a full OS, while VM does. Second, the memory allocation is more flexible for containers than for VMs. By comparing Figure \ref{32docker_mem} and Figure \ref{32vm_mem}, it shows that on one hand, the amount of memory allocated to a container is very small at the beginning, and then increases based on the demands of the applications in the container. However, a VM occupies more memory at the beginning. On the other hand, a container releases its memory after it finishes its workload, while a VM still holds the memory even after it becomes idle.

%\subsection{Isolation}

%\subsection{Resource Sensitivity}
%File system, I/O mode
%\subsection{Migration}
%\subsection{Container Configuration}
%1. Running with different storage configurations. Q: besides HDFS, does a Spark job write anything to the local file system? Maybe logs? 
%2. JVM running inside container

%\subsection{Energy consumption}

\section{Conclusions and Future Work} \label{conclusion}
We have presented an in-depth experimental study between virtual machine and Docker container in Spark big data environment. Our extensive measurement study shows a number of interesting observations: (i) Dockers container is more convenient than virtual machine for system administrators both in deployment and bootup stages. (ii) with different big data workloads, Dockers container shows much better scalability than virtual machines. (iii) with the same workload, Dockers container achieves higher CPU and memory utilization. We conjecture that our study results can help practitioners and researchers to make more informed decisions on tuning their cloud environment and configuring their big data applications, so as to achieve better performance and higher resources utilization. Many more aspects of container and virtual machine environment will be explored as our future work. For instance, how efficient is the container migration versus the virtual machine migration, and what impact will such migration mechanisms have on the performance of big data applications running in the cloud. Another direction is to compare the design and organization of the image file between containers and virtual machines, and investigate its impact on the applications performance.
%\section*{Acknowledgment}

\bibliography{docker_cloud}
\bibliographystyle{ieee}

\end{document}